%% file: main.tex
\documentclass[]{vgtc}                %
\ifpdf
  \pdfoutput=1\relax                   
  \usepackage{graphicx}                
  \DeclareGraphicsExtensions{.pdf,.png,.jpg,.jpeg} 
\else
  \usepackage{graphicx}                
  \DeclareGraphicsExtensions{.eps}     
\fi%

\graphicspath{{figures/}{pictures/}{images/}{./}} 

\usepackage{microtype}                 
\PassOptionsToPackage{warn}{textcomp}  
\usepackage{textcomp}                  
\usepackage{mathptmx}                  
\usepackage{times}                     
\usepackage{cite}                      
\usepackage{tabu}                      
\usepackage{booktabs}                  
\usepackage{multirow}
\usepackage{array}
\usepackage{colortbl}
\usepackage{xcolor}
\usepackage{tikz}
\usepackage{bm}
\usepackage{amssymb}

\newcommand\hl[1]{%
  \bgroup
  \hskip0pt\color{red!80!black}%
  #1%
  \egroup
}%

\onlineid{0}

\vgtccategory{Research}
\vgtcpapertype{please specify}

\title{Enabling Longitudinal Exploratory Analysis of Clinical COVID Data}

\author{
David Borland\thanks{email: borland@renci.org}\\ \scriptsize{RENCI, UNC-Chapel Hill} \and 
Irena Brain \thanks{e-mail: irenab@live.unc.edu}\\ \scriptsize UNC-Chapel Hill \and
Karamarie Fecho \thanks{e-mail: kfecho@renci.org}\\ \scriptsize RENCI, UNC-Chapel Hill \and
Emily Pfaff \thanks{e-mail: epfaff@email.unc.edu}\\ \scriptsize UNC-Chapel Hill \and
Hao Xu \thanks{e-mail: xuhao@renci.org}\\ \scriptsize RENCI, UNC-Chapel Hill \and
James Champion \thanks{e-mail: champioj@email.unc.edu}\\ \scriptsize UNC-Chapel Hill \and
Chris Bizon\thanks{e-mail: bizon@renci.org}\\ \scriptsize RENCI, UNC-Chapel Hill  \and
David Gotz\thanks{e-mail: gotz@unc.edu}\\ \scriptsize UNC-Chapel Hill
}

\shortauthortitle{Borland \MakeLowercase{\textit{et al.}}: Enabling Longitudinal Exploratory Analysis of Clinical COVID Data}

\abstract{
As the COVID-19 pandemic continues to impact the world, data is being gathered and analyzed to better understand the disease. Recognizing the potential for visual analytics technologies to support 
exploratory analysis and hypothesis generation from longitudinal clinical data, a team of collaborators worked to apply existing event sequence visual analytics technologies to a longitudinal clinical data from a cohort of 998 patients with high rates of COVID-19 infection.
This paper describes the initial steps toward this goal, including: (1) the data transformation and processing work required to prepare the data for visual analysis, (2) initial findings and observations, and (3) qualitative feedback and lessons learned which highlight key features as well as limitations to address in future work.
} 

\keywords{Visual analytics, temporal event sequence visualization, human-computer interaction, medical informatics, COVID-19}

\CCScatlist{ 
 \CCScat{Human-centered computing}{Visualization}{Visualization systems and tools};
 \CCScat{Human-centered computing}{Visualization}{Visualization application domains}{Visual analytics}
}

\teaser{
  \centering
  \includegraphics[width=1.0\linewidth]{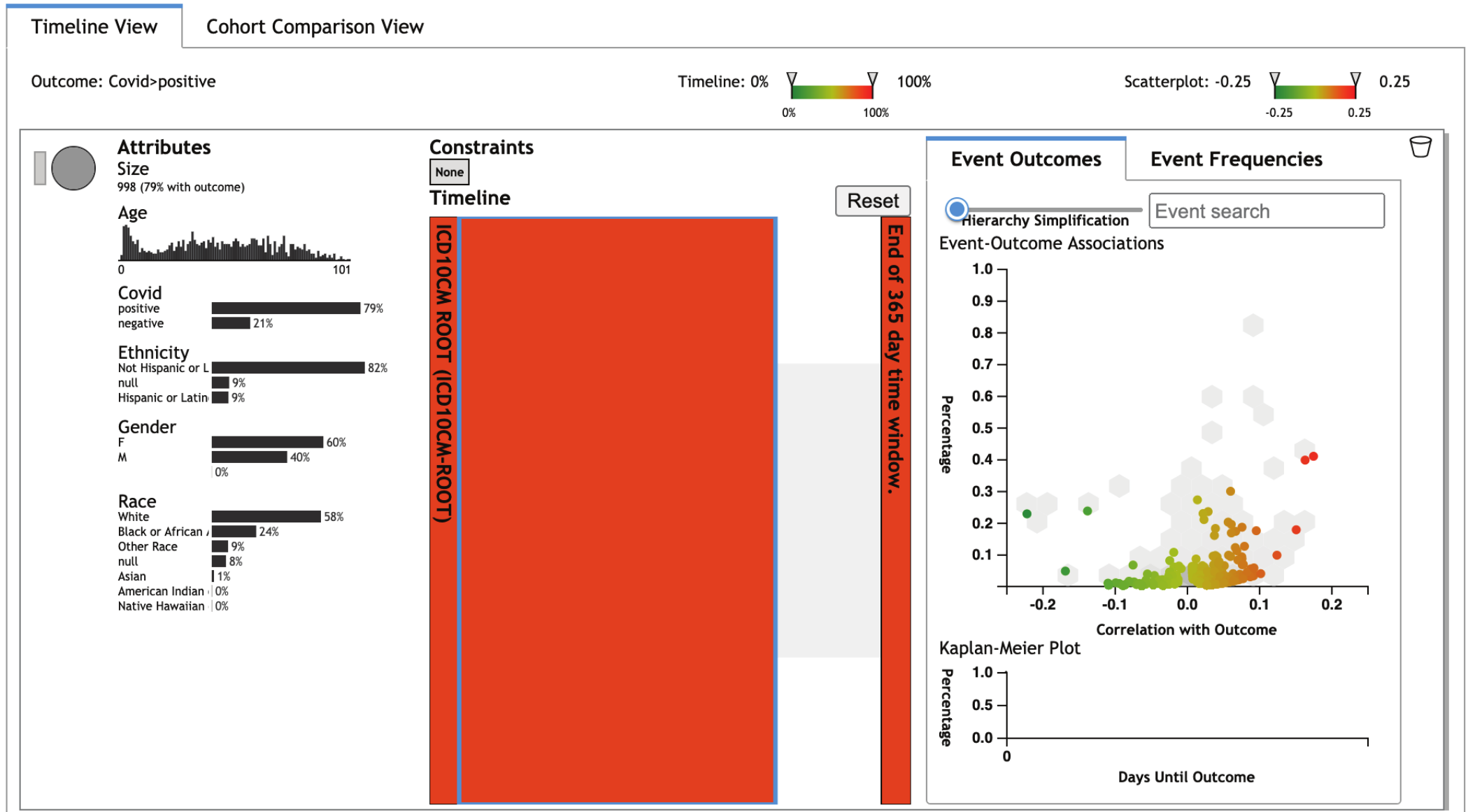}
  \caption{\label{analysis:main} The Cadence visual analytics system in use to examine data from a cohort of 998 patients from early in the COVID-19 pandemic. Because no official COVID diagnosis was available in the standard coding systems at that point in the pandemic, a temporal query of ``First ICD-10 Diagnosis'' followed by a one year time window was issued to select all patients and all events in the available data. As the visualization shows, the patients were 79\% COVID-positive (based on a standardized COVID phenotype), 60\% female, and represented a wide variety of ages, from infants to elderly adults.  An exploration of potential risk factors based on this cohort is presented within this paper.}
 	\label{fig:teaser}
}

\begin{document}

\input{sections/intro}
\input{sections/related}
\input{sections/methods}

\input{sections/discussion}

\input{sections/conclusion}

\acknowledgments{
The research reported in this article was supported in part by a grant from the United States National Science Foundation (\#1704018). Partial support was also provided by the United States National Institutes of Health (NCATS OT2TR003430,  NCATS UL1TR002489,  NCATS UL1TR002489-03S4, and NCATS OT3TR002020).}

\bibliographystyle{abbrv-doi}

\bibliography{main}

\onecolumn
\appendix

\section{Additional Figures}

\begin{figure}[h]
\centering
  \includegraphics[width=1.0\linewidth]{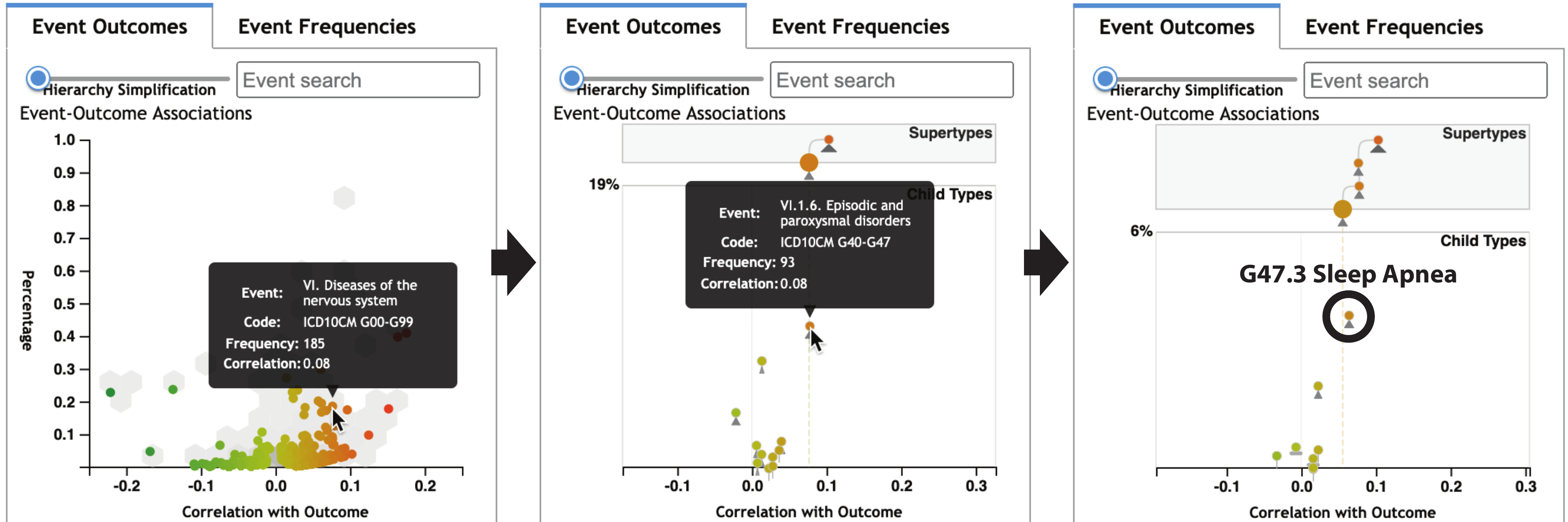}
  \caption{Diagnoses for diseases of the nervous system showed positive correlation with COVID-19-positive status. Drilling down via the interactive scatterplot, it was found that Sleep Apnea was the most common diagnosis code within that category and exhibited a positive correlation.}
  \label{fig:sleepapnea}
\end{figure}

\begin{figure}[h]
\centering
  \includegraphics[width=0.7\linewidth]{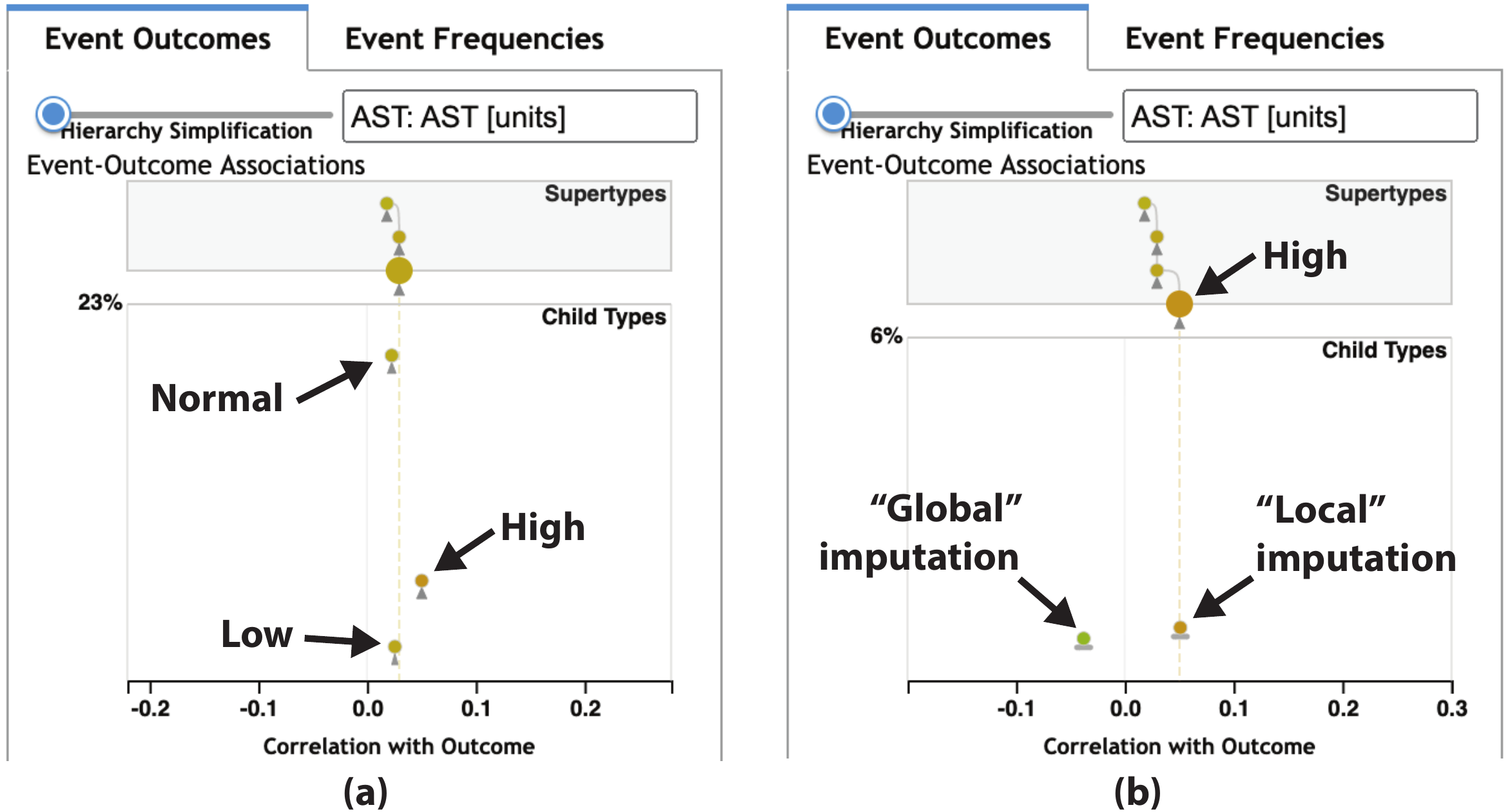}
  \caption{Several lab tests that were thought to be associated with COVID-19 infection proved to have weaker association than expected in the study cohort.  (a) This example screenshot shows that high AST tests had a stronger positive correlation with COVID than normal or low tests, but the effect was very weak. Meanwhile, the scenting feature of Cadence (the gray triangular glyph underneath the High dot in the plot) suggests that variation exists between subtypes of High AST. Clicking on the High dot brings up (b) a visualization of the High result's subtypes which correspond to different methods of imputation. The plot shows that local imputation methods perform well in that the locally imputed High lab values have similar correlation statistics to the High test results found in the raw data.  However, global imputation appears to have performed poorly as indicated by the large difference in correlation to COVID-positive patients. In fact, the sign of the correlation is reversed. This suggests a potential flaw in the global imputation approach, a key insight for analysts.} 
  \label{fig:lab}
\end{figure}

\end{document}

%% file: sections/intro.tex
\firstsection{Introduction}

\maketitle

Beginning in late 2019, the emergence of the SARS-CoV-2 virus and its corresponding disease COVID-19 triggered a quickly spreading global pandemic. The crisis has caused health and economic harms to billions of people around the world, and resulted in millions of deaths in less than two years of existence. As of this writing, novel variants are emerging and continue to ravage unvaccinated populations \cite{callaway2021could}. As a result, and despite the development of effective vaccines, it appears that the virus is becoming endemic \cite{torjesen2021covid}. 

This ongoing health crisis has led to significant investments in research to better understand the nature of the SARS-CoV-2 virus, the novel disease it causes, risk factors that relate to severe outcomes, and the efficacy of potential treatments. One major initiative sponsored by the National Institutes of Health in the United States is the National COVID Cohort Collaborative (N3C) \cite{ncats_n3c, bennett_national_2021}. N3C is a multi-institutional effort to collect and share longitudinal clinical data in support of urgent research related to the pandemic.

Given the novelty of the COVID-19 disease--including many unknowns surrounding long and short-term symptoms, risk factors, and response to treatments--there is great interest among researchers in finding effective tools that can facilitate exploratory analysis and hypothesis generation based on complex longitudinal clinical data. 
These requirements align well with the general goals of visual analytics technologies \cite{thomas_illuminating_2005,gotz_data-driven_2016}, and fit particularly closely with the capabilities of certain visual analytics tools developed for understanding event sequence data \cite{arxiv_2020_guo}. 

Motivated by this alignment in analytic requirements and system capabilities, a collaborative project was initiated to apply Cadence \cite{gotz_visual_2020}, an existing prototype visual analytics tool for event sequence analysis, to COVID-centric clinical data gathered at UNC Health using the N3C phenotype definition. This paper describes key aspects of this project, including: (1) the data transformation and preparation work required to ready the COVID cohort data for visual analysis, (2) initial findings and observations from a visual analysis of the data, and (3) qualitative feedback and lessons learned which include observations that motivate future research opportunities.

%% file: sections/related.tex
\section{Related Work}

This section provides an overview of two key areas of research that provide a context for the work presented in this paper: applications of visualization to support the response to the COVID-19 pandemic, and visual analytics techniques for event sequence analysis.

\subsection{Visualization and COVID-19}

The value of visualization for health-focused applications has been well-studied \cite{gotz_data-driven_2016}, including both clinical use cases \cite{west_innovative_2015} and public health applications \cite{preim_survey_2020}.  Accordingly, visualization has played a key role in the ongoing response to the COVID-19 pandemic. Examples include visualizations for managing the rise in telehealth activity \cite{dixit_rapid_2020}, surveillance tools for tracking infection prevalence geographically and over time \cite{dixon_leveraging_2021}, and fatality management tools for managing the worst of the pandemic's impacts \cite{kaul_rapidly_2020}. One common theme in these efforts is the need for rapid development and innovation to meet the needs of a quickly changing environment. Our work follows a similar rapid development process. However, in contrast to the COVID-related articles cited above, which primarily focus on situational awareness and crisis management for on-the-ground response, the work presented in this paper aims to support COVID research activities for analysts working to better understand the nature of the disease.

\subsection{Event Sequence Analysis}

Event sequence data has been the focus of a large number of visual analytics research efforts \cite{arxiv_2020_guo}. From techniques for individual event sequences (e.g., \cite{plaisant_lifelines:_1998}) to those for large collections of sequences (e.g., \cite{wongsuphasawat_lifeflow:_2011,wongsuphasawat_outflow:_2011}), many of these technologies have been developed with medical data analysis as a primary application. More recent work has focused on solving scalability challenges for these types of visual analytics tools \cite{du_coping_2017}. Such challenges are especially pronounced in medical data, such as the extensive number of event types that arise from large medical coding systems \cite{gotz_decisionflow:_2014,gotz_visual_2020}.

Reflecting the need for rapid progress, the work reported in this paper leverages Cadence \cite{zhang_dynamic_2019}, an existing visual analytics platform. Cadence employs a scalable visual design similar to the earlier DecisionFlow design \cite{gotz_decisionflow:_2014}, but adds selection bias quantification tools \cite{borland_selection_2020,borland_selection-bias-corrected_2021} and data aggregation and navigation features for hierarchical event types such as those found in coded medical data \cite{gotz_visual_2020}.  This hierarchical aggregation and navigation feature proved especially useful for the work reported in this paper because, beyond the normal benefits for high-dimensional health data, it provided a way to allow users to optionally distinguish between different imputation methods during as part of their analysis.

%% file: sections/methods.tex
\section{Methods and Findings}
\label{sec:methods}

To support exploratory analysis and hypothesis generation by researchers investigating the clinical nature of COVID-19, we applied the Cadence visual analytics system to longitudinal medical data from UNC Health that was collected using the N3C phenotype definition in collaboration with the Integrated Clinical and Environmental Exposures Services (ICEES) \cite{fecho_novel_2019,ahalt_clinical_2019} team at UNC-Chapel Hill. 

ICEES is an open regulatory-compliant service that exposes sensitive clinical data (e.g., EHR data, survey data, research data) that have been integrated at the patient level with a variety of public environmental exposures data (e.g., airborne pollutants, major roadways/highways, socio-economic factors, landfills, CAFOs).  ICEES provides disease-agnostic capabilities for data access and analysis, and was recently adapted to support research activities surrounding coronavirus infections.

This section details the key steps required to apply the Cadence visual analytics system in this context, and describes an initial set of interesting findings. This work represents the first steps towards the ultimate aim of fully integrating Cadence capabilities within the larger ICEES platform for longitudinal analysis of large scale clinical data.

\begin{figure}
\centering
  \includegraphics[width=0.9\linewidth]{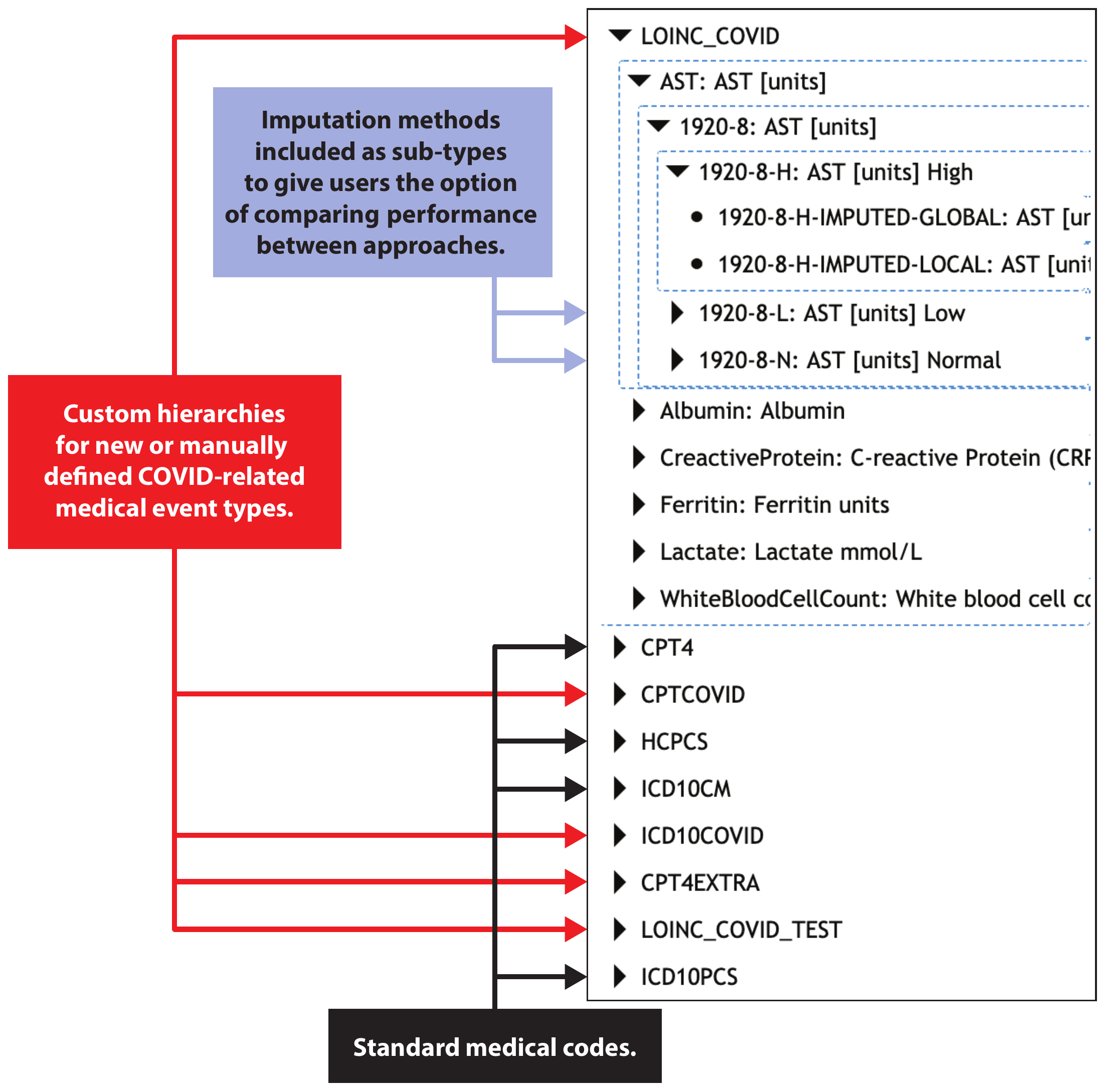}
  \caption{The type hierarchy for this project includes a combination of standard coding systems, custom hierarchies for new COVID-specific concepts, and representations of different imputation methods.}
  \label{fig:types}
\end{figure}

\subsection{Data Preparation}

Data for an initial set of 999 patients were obtained from the N3C team at UNC-Chapel Hill.  The data included patient demographic attributes (including ethnicity, gender, race, and age), diagnoses, procedures, and laboratory tests. One patient was excluded because there were zero diagnoses present in that patient's data, resulting in an  overall cohort of 998 patients. The cohort was 60\% female and included patients of all ages from infants to elderly adults.  These figures are visible in the Attributes sidebar on the left side of Figure~\ref{fig:teaser}.  The data were represented using standard medical coding systems (e.g., ICD-10, CPT4, and LOINC), and included several newly added codes created specifically for COVID-related concepts. 

The data also contained labels for each patient identifying their COVID-19 status (COVID positive or COVID negative).  Because the data for this cohort was captured early in the pandemic, few patients had actual COVID tests. Instead, labels were assigned to each patient based on a specific set of criteria regarding combinations of diagnoses and lab test results.  In all, 79\% of the cohort was labeled as COVID positive.

Data was provided in a set of JSON-encoded FHIR format files \cite{json_fhir} and transformed into the Cadence data model.  This model represents time-specific events as $<ID,Date,EventType>$ triples where the $EventType$ has both a class and a code component (e.g., $<$ICD-10,B34.2$>$ to represent a coronavirus infection using the ICD-10 coding system).  As part of this process, all diagnosis and procedure data was included.  However, only LOINC data specifically related to COVID was included in the transformation from FHIR to Cadence formats.  This includes LOINC-coded data about COVID tests (for those that had access, which was limited at the time) and other lab tests thought to be associated with COVID-related outcomes based on published literature \cite{bennett_national_2021}.  Other LOINC lab tests were excluded from the analysis due to the manual work required to include these values (see the next section for more details).

\subsection{Data Imputation and Categorization}
\label{sec:imputation}

Unlike diagnoses or procedures, which are represented as categorical events (e.g., a pneumonia diagnosis made on a specific date), lab tests are typically associated with attributes.  For example, a white blood cell count test would include not only the fact that the test occurred, but also the value of the result.  To fit this type of data into the triplet format used by Cadence, the lab results needed to be converted into categorical values. To achieve this goal, we used reference ranges for the test results to map observed numerical values to either High, Normal, or Low categories.  Then, within the data, we created three types for each lab test (e.g., White-Blood-Cell-HIGH, White-Blood-Cell-NORMAL, White-Blood-Cell-LOW). 

This process worked predictably and reliably when reference range data was available in the original FHIR data. However, we found information about High or Low ranges were often missing.  Moreover, the specific ranges used to determine High or Low values can vary from patient to patient in complex ways based on their medical condition.  

We therefore applied a two-phase imputation algorithm to fill in missing high or low reference range information.   First, in a \emph{local imputation phase}, if a given occurrence of a lab test was missing a reference range, we looked at the same patient's data to see if there were any records of the same test being performed on different dates that included a reference range. If so, we applied the reference range most frequently used for that patient on other dates to the lab test that was missing a reference range. Second, in a \emph{global imputation phase}, if a given occurrence of a lab test was missing a reference range and all occurrences of that lab test for the same patient were also missing reference ranges, we applied the most frequent reference range from the overall population to that lab test.   

\subsection{Event Type Metadata}

Cadence leverages event type hierarchies to manage complexity by performing aggregation of similar event types.  The system recommends informative levels of aggregation while enabling users to manually move up or down the hierarchy of event types as desired during analysis.  Standard medical codes are used as much as possible by leveraging the open-source Athena vocabulary database maintained by OHDSI \cite{athena}.  However, the advent of COVID has led to the introduction of a number of extremely new codes. One key feature of Cadence is that it allows manually defined type hierarchies to supplement those in Athena, and this feature was used to add support for any new codes found in the data that were not present in Athena.  These event types are shown in Figure~\ref{fig:types}.

The manually defined type hierarchy feature of Cadence was also critical for maintaining transparency for imputed event types.  The imputation process described in Section~\ref{sec:imputation} produces a High, Normal, or Low label for all lab tests. It was important to ensure that analysts could both: (1) treat all High (or Normal, or Low) values the same regardless of imputation method to maximize statistical power, and (2) distinguish between imputation methods to identify batch effects for validation of the various approaches. This was accomplished by inserting multiple event types into the hierarchy for each categorized lab test result, one for each imputation method.  
This can be seen in the user interface snapshot shown in Figure~\ref{fig:types} for the AST lab test.  The expanded hierarchy shows three subtypes of AST tests: High, Low, and Normal.  The expanded High value section shows subtypes which can be used to represent High values obtained via the two types of imputation.  This approach enables comparisons of different imputation methods as shown in Figure~\ref{fig:lab}(b).

\subsection{Findings}

Once the data processing stage was complete, the data were loaded into the Cadence event sequence visual analysis platform for analysis.  The data were initially explored by developers of the Cadence system. Then two interactive meetings were held with members of the collaborating ICEES team at UNC-Chapel Hill. The first meeting included two developers and five other full-time research staff collaborators, including three PhD-level researchers involved in medical data analysis. The second meeting included two developers and one PhD-level researcher in medical data analysis. During these meetings, the visualization was displayed to all attendees using Zoom screen sharing and jointly analyzed as we discussed the findings.

The first step in using Cadence was to issue a query. Event sequence visualization tools, including Cadence, generally aim to temporally align event sequences by a sentinel event (e.g., first diagnosis of COVID-19) to see patterns in how patients evolve before or after the alignment point.  However, in the early days of the pandemic, patients were not often formally diagnosed with COVID-19. The disease was novel and not yet represented as a coded diagnosis within the EHR systems. Therefore, data were retrieved by issuing a temporal query that asked for one year of medical data following the first occurrence of any ICD-10 code.  The generic nature of this query ensured that all 998 patients in the cohort matched the query, and the one year duration of the time window was sufficiently long to ensure that all available data was included in the analysis. The COVID-19 label value for these patients was used as the outcome with a positive COVID-19 value representing a negative outcome for the purposes of analysis.

The result from this initial query is shown in Figure~\ref{fig:teaser}. The Attributes column provides a summary of key non-temporal patient data including gender, ethnicity, race, and age distributions. In addition, the summary view showed that 79\% of patients were labeled as COVID-positive.

Analysts were next drawn to the event scatterplot on the right side of the Cadence interface. Each circle in this plot represents a specific event type or higher-level event type group (e.g., a range of closely related ICD-10 codes) as determined by a dynamic hierarchical aggregation algorithm that aims to maximize informativeness of the visualized representation with respect to the outcome (in this case, a patient's COVID-19 label) \cite{gotz_visual_2020}. The x-axis represents the event type's correlation with the outcome, and the y-axis represents the proportion of the population that exhibits the corresponding event type. Given this encoding, the most ``interesting'' events tend to be those on the left/right extremes (higher correlation) and toward the top of the scatterplot (high frequency).

Via interactive probing of this plot, the team was able to quickly confirm that well-publicised COVID-19 symptoms were indeed among the most strongly associated factors with a positive COVID-19 label.  Factors standing out at the periphery of the scatterplot included cough, fever, pneumonia, and a range of respiratory issues including general difficulty in breathing. The very quick identification and confirmation of these well-known symptoms raised the team's confidence in the system's analysis capabilities.  

Similarly, a strong negative indicator for COVID-19 was a negative COVID test.  These tests were rare early in the pandemic, and some patients who tested negative later tested positive.  However, a negative test was by far the strongest negatively associated factor.

Not all findings were confirmatory, however, with the visualization highlighting some surprising discoveries.  For example, there was a strong negative association between an ``exposure to communicable disease'' ICD-10 diagnosis and a positive COVID-19 status.  Given the focus on contact tracing and our understanding of how disease spreads, this is opposite from what might be initially expected.  However, surprisingly, the statistics for this diagnosis placed it quite close to a negative COVID-19 test in the scatterplot.  A discussion within the team ensued to interpret this finding, and the leading hypothesis was that early in the pandemic, fearful patients with no symptoms were arriving for medical care because of potential exposure as their only risk factor. Without observable symptoms or the availability of confirmatory tests, these patients were not considered COVID-positive.

Similarly, the team was somewhat surprised to see a lack of strong support for some lab tests that had been expected to be predictive indicators for positive COVID cases: AST, Albumin, C-Reactive Protein, Ferritin, Lactate, and White Blood Cell Count \cite{bennett_national_2021}. Abnormal tests did show some correlation with positive COVID cases, but the strength of the effect was very small.  For example, Figure~\ref{fig:lab}(a) shows the data for AST tests in our cohort. This screenshot from Cadence shows that the High AST test results did have a stronger correlation than Normal or Low tests, but all were close to zero. 

This could be caused in part by the imbalance in our cohort (70\% COVID-positive) as well as the fact that these patients were from early in the pandemic before standards of care were developed.  However, we can also see another potential problem from the visualization.  The gray triangle beneath the High circle in Figure~\ref{fig:lab}(a) is wide.  This is a scenting clue provided by Cadence that variation exists within children event types. 

Clicking on the High circle brought us to the view in Figure~\ref{fig:lab}(b).  Here we can see that not all High lab tests are the same.  Locally imputed high AST test results were statistically similar to those found in the raw data. Globally imputed values, however, were quite different. This suggests a potential problem with the global imputation algorithm which may be obscuring more meaningful relationships between labs and a patient's COVID status.  This shows the critical value in ensuring that analysts can distinguish between observed data, imputed values, and the different methods used for imputation.

Other interesting findings included a surprising-to-the-team link between sleep apnea and COVID-19 status. As shown in Figure~\ref{fig:sleepapnea}, the general class of diseases of the nervous system (ICD codes in range G00-G99) had a positive correlation with COVID-19.  Drilling down via the visualization, we found sleep apnea to be the predominant diagnoses within this category.  This was a new discovery for those of us in the meeting, but a subsequent literature search demonstrated that this finding is in fact in line with recently published research \cite{strausz_sleep_2021}. 
In a similar way, we found relatively strong links between various hyperlipidemia diagnoses and COVID-19 status.  This too was confirmed by a subsequent literature search as an exacerbating factor \cite{wang_role_2021, butt_association_2020}.

%% file: sections/discussion.tex
\section{Discussion} \label{disc}

Beyond the specific findings about the data, our analyses and the two interactive team meetings described in Section \ref{sec:methods} yielded several insights about both the utility of various event sequence analysis features and open challenges that would be valuable to address in future work.

For example, the collaborators found the attribute summaries useful to ground the work.  ``That's handy, the demographics.'' Commenting on the gender breakdown, ``that's about the hospital breakdown.'' This along with confirmation of known symptoms increased user confidence in the system.

The ability to browse visually up and down the event type hierarchy as part of the visualization was seen to be especially valuable. A collaborator commented ``that's a nice feature I think, often you care about the higher level'' aggregation. In addition, it enabled users to search for generic terms (e.g., `vent' for `ventilator') and then browse from that point to find a specific code or code group. Relatedly, the ability to leverage the type hierarchy to distinguish between observed values and different types of imputed values proved extremely valuable. As described in Section~\ref{sec:methods} with the AST lab test example, this approach provided essential transparency as to the batch effects between different methods of imputation.

Beyond these useful capabilities, the surprise findings were often most exciting for users. Remarking on the previously described sleep apnea observation, one user exclaimed ``Oh, OK. That's interesting!'' Users would then immediately begin thinking about the potential validity of the discovery. Showing her internal thought process, a user commented about the sleep apnea discovery by stating ``it makes sense. It's a big risk factor for anesthesiology too.''

In terms of limitations, there were several that would benefit from additional research.  One key constraint was the requirement that events with attribute values (e.g. lab tests) be mapped to categorical values (e.g., High, Normal, Low).  This forces an analyst to impose arbitrary thresholds. This can obscure potentially valuable information such as trending in labs performed multiple times, or differences in thresholds for different patients. There has been some past research exploring events with attributes \cite{cappers2017exploring} but continued research is needed.   Similarly, the pre-defined type hierarchy imposes constraints on the units of analysis.  Support for more dynamic and flexible grouping would bring these tools closer to allowing arbitrary value sets \cite{valuesets} during analysis.

Finally, one of our collaborators mentioned that we could ``do cool stuff with other outcomes and a better dataset.'' For future work, we aim to apply Cadence to a larger 100,000 patient cohort from from UNC Health that has a more balanced case/control population and which includes samples from later in the pandemic.  This will enable us, as requested by our collaborators, to look more closely at different outcomes, such as mortality, ventilation, or other indicators of severe disease.

%% file: sections/conclusion.tex
\section{Conclusion}

A collaborative effort was undertaken to adapt and apply existing visual analytics technologies to support exploratory analysis and hypothesis generation from from UNC Health data gathered using the COVID phenotype definition developed by the National COVID Cohort Collaborative (N3C). This paper describes the initial steps toward this goal, including: (1) the data transformation and preparation work required to prepare the data for visual analysis, (2) initial findings and observations, and (3) qualitative feedback and lessons learned about the visual analytics system which highlight both useful features and limitations to address in future work.

%% file: main.bbl
\begin{thebibliography}{10}

\bibitem{ahalt_clinical_2019}
S.~C. Ahalt, C.~G. Chute, K.~Fecho, G.~Glusman, J.~Hadlock, C.~O. Taylor, E.~R.
  Pfaff, P.~N. Robinson, H.~Solbrig, C.~Ta, N.~Tatonetti, and C.~Weng.
\newblock Clinical {Data}: {Sources} and {Types}, {Regulatory} {Constraints},
  {Applications}.
\newblock {\em Clinical and Translational Science}, 12(4):329--333, 2019. doi:
  {{%
10\hspace{.1pt}\discretionary{.}{%
}{.}\hspace{.4pt}1111\discretionary{/}{%
}{/}cts\hspace{.1pt}\discretionary{.}{%
}{.}\hspace{.4pt}12638}}


\bibitem{bennett_national_2021}
T.~D. Bennett, R.~A. Moffitt, J.~G. Hajagos, B.~Amor, A.~Anand, M.~M. Bissell,
  K.~R. Bradwell, C.~Bremer, J.~B. Byrd, A.~Denham, P.~E. DeWitt, D.~Gabriel,
  B.~T. Garibaldi, A.~T. Girvin, J.~Guinney, E.~L. Hill, S.~S. Hong,
  H.~Jimenez, R.~Kavuluru, K.~Kostka, H.~P. Lehmann, E.~Levitt, S.~K.
  Mallipattu, A.~Manna, J.~A. McMurry, M.~Morris, J.~Muschelli, A.~J. Neumann,
  M.~B. Palchuk, E.~R. Pfaff, Z.~Qian, N.~Qureshi, S.~Russell, H.~Spratt,
  A.~Walden, A.~E. Williams, J.~T. Wooldridge, Y.~J. Yoo, X.~T. Zhang, R.~L.
  Zhu, C.~P. Austin, J.~H. Saltz, K.~R. Gersing, M.~A. Haendel, and C.~G.
  Chute.
\newblock The {National} {COVID} {Cohort} {Collaborative}: {Clinical}
  {Characterization} and {Early} {Severity} {Prediction}.
\newblock {\em medRxiv}, p. 2021.01.12.21249511, Jan. 2021. doi: {{%
10\hspace{.1pt}\discretionary{.}{%
}{.}\hspace{.4pt}1101\discretionary{/}{%
}{/}2021\hspace{.1pt}\discretionary{.}{%
}{.}\hspace{.4pt}01\hspace{.1pt}\discretionary{.}{%
}{.}\hspace{.4pt}12\hspace{.1pt}\discretionary{.}{%
}{.}\hspace{.4pt}21249511}}


\bibitem{borland_selection_2020}
D.~Borland, W.~Wang, J.~Zhang, J.~Shrestha, and D.~Gotz.
\newblock Selection {Bias} {Tracking} and {Detailed} {Subset} {Comparison} for
  {High}-{Dimensional} {Data}.
\newblock {\em IEEE Transactions on Visualization and Computer Graphics},
  26(1), 2020.

\bibitem{borland_selection-bias-corrected_2021}
D.~Borland, J.~Zhang, S.~Kaul, and D.~Gotz.
\newblock Selection-{Bias}-{Corrected} {Visualization} via {Dynamic}
  {Reweighting}.
\newblock {\em IEEE Transactions on Visualization and Computer Graphics},
  27(2):1481--1491, 2021. doi: {{%
10\hspace{.1pt}\discretionary{.}{%
}{.}\hspace{.4pt}1109\discretionary{/}{%
}{/}TVCG\hspace{.1pt}\discretionary{.}{%
}{.}\hspace{.4pt}2020\hspace{.1pt}\discretionary{.}{%
}{.}\hspace{.4pt}3030455}}


\bibitem{butt_association_2020}
J.~H. Butt, T.~A. Gerds, M.~Schou, K.~Kragholm, M.~Phelps, E.~Havers-Borgersen,
  A.~Yafasova, G.~H. Gislason, C.~Torp-Pedersen, L.~Køber, and E.~L. Fosbøl.
\newblock Association between statin use and outcomes in patients with
  coronavirus disease 2019 ({COVID}-19): a nationwide cohort study.
\newblock {\em BMJ Open}, 10(12):e044421, Dec. 2020. doi: {{%
10\hspace{.1pt}\discretionary{.}{%
}{.}\hspace{.4pt}1136\discretionary{/}{%
}{/}bmjopen\discretionary{%
}{-}{-}2020\discretionary{%
}{-}{-}044421}}


\bibitem{callaway2021could}
E.~Callaway.
\newblock Could new covid variants undermine vaccines? labs scramble to find
  out.
\newblock {\em Nature}, 589(7841):177--178, 2021.

\bibitem{cappers2017exploring}
B.~C. Cappers and J.~J. van Wijk.
\newblock Exploring multivariate event sequences using rules, aggregations, and
  selections.
\newblock {\em IEEE transactions on visualization and computer graphics},
  24(1):532--541, 2017.

\bibitem{dixit_rapid_2020}
R.~A. Dixit, S.~Hurst, K.~T. Adams, C.~Boxley, K.~Lysen-Hendershot, S.~S.
  Bennett, E.~Booker, and R.~M. Ratwani.
\newblock Rapid development of visualization dashboards to enhance situation
  awareness of {COVID}-19 telehealth initiatives at a multihospital healthcare
  system.
\newblock {\em Journal of the American Medical Informatics Association},
  27(9):1456--1461, Sept. 2020. doi: {{%
10\hspace{.1pt}\discretionary{.}{%
}{.}\hspace{.4pt}1093\discretionary{/}{%
}{/}jamia\discretionary{/}{%
}{/}ocaa161}}


\bibitem{dixon_leveraging_2021}
B.~E. Dixon, S.~J. Grannis, C.~McAndrews, A.~A. Broyles, W.~Mikels-Carrasco,
  A.~Wiensch, J.~L. Williams, U.~Tachinardi, and P.~J. Embi.
\newblock Leveraging data visualization and a statewide health information
  exchange to support {COVID}-19 surveillance and response: {Application} of
  public health informatics.
\newblock {\em Journal of the American Medical Informatics Association},
  28(7):1363--1373, July 2021. doi: {{%
10\hspace{.1pt}\discretionary{.}{%
}{.}\hspace{.4pt}1093\discretionary{/}{%
}{/}jamia\discretionary{/}{%
}{/}ocab004}}


\bibitem{du_coping_2017}
F.~Du, B.~Shneiderman, C.~Plaisant, S.~Malik, and A.~Perer.
\newblock Coping with {Volume} and {Variety} in {Temporal} {Event} {Sequences}:
  {Strategies} for {Sharpening} {Analytic} {Focus}.
\newblock {\em IEEE Transactions on Visualization and Computer Graphics},
  23(6):1636--1649, June 2017. doi: {{%
10\hspace{.1pt}\discretionary{.}{%
}{.}\hspace{.4pt}1109\discretionary{/}{%
}{/}TVCG\hspace{.1pt}\discretionary{.}{%
}{.}\hspace{.4pt}2016\hspace{.1pt}\discretionary{.}{%
}{.}\hspace{.4pt}2539960}}


\bibitem{fecho_novel_2019}
K.~Fecho, E.~Pfaff, H.~Xu, J.~Champion, S.~Cox, L.~Stillwell, D.~B. Peden,
  C.~Bizon, A.~Krishnamurthy, A.~Tropsha, and S.~C. Ahalt.
\newblock A novel approach for exposing and sharing clinical data: the
  {Translator} {Integrated} {Clinical} and {Environmental} {Exposures}
  {Service}.
\newblock {\em Journal of the American Medical Informatics Association},
  26(10):1064--1073, Oct. 2019. doi: {{%
10\hspace{.1pt}\discretionary{.}{%
}{.}\hspace{.4pt}1093\discretionary{/}{%
}{/}jamia\discretionary{/}{%
}{/}ocz042}}


\bibitem{gotz_data-driven_2016}
D.~Gotz and D.~Borland.
\newblock Data-{Driven} {Healthcare}: {Challenges} and {Opportunities} for
  {Interactive} {Visualization}.
\newblock {\em IEEE Computer Graphics and Applications}, 36(3):90--96, May
  2016. doi: {{%
10\hspace{.1pt}\discretionary{.}{%
}{.}\hspace{.4pt}1109\discretionary{/}{%
}{/}MCG\hspace{.1pt}\discretionary{.}{%
}{.}\hspace{.4pt}2016\hspace{.1pt}\discretionary{.}{%
}{.}\hspace{.4pt}59}}


\bibitem{gotz_decisionflow:_2014}
D.~Gotz and H.~Stavropoulos.
\newblock {DecisionFlow}: {Visual} {Analytics} for {High}-{Dimensional}
  {Temporal} {Event} {Sequence} {Data}.
\newblock {\em IEEE Transactions on Visualization and Computer Graphics},
  20(12):1783--1792, 2014. doi: {{%
10\hspace{.1pt}\discretionary{.}{%
}{.}\hspace{.4pt}1109\discretionary{/}{%
}{/}TVCG\hspace{.1pt}\discretionary{.}{%
}{.}\hspace{.4pt}2014\hspace{.1pt}\discretionary{.}{%
}{.}\hspace{.4pt}2346682}}


\bibitem{gotz_visual_2020}
D.~Gotz, J.~Zhang, W.~Wang, J.~Shrestha, and D.~Borland.
\newblock Visual {Analysis} of {High}-{Dimensional} {Event} {Sequence} {Data}
  via {Dynamic} {Hierarchical} {Aggregation}.
\newblock {\em IEEE Transactions on Visualization and Computer Graphics},
  26(1), 2020.

\bibitem{arxiv_2020_guo}
Y.~Guo, S.~Guo, Z.~Jin, S.~Kaul, D.~Gotz, and N.~Cao.
\newblock A survey on visual analysis of event sequence data.
\newblock {\em IEEE Transactions on Visualization and Computer Graphics}, Early
  Access, 2021. doi: {{%
10\hspace{.1pt}\discretionary{.}{%
}{.}\hspace{.4pt}1109\discretionary{/}{%
}{/}TVCG\hspace{.1pt}\discretionary{.}{%
}{.}\hspace{.4pt}2021\hspace{.1pt}\discretionary{.}{%
}{.}\hspace{.4pt}3100413}}


\bibitem{json_fhir}
{HL7 FHIR Community}.
\newblock {JSON - FHIR v4.0.1}.
\newblock https://www.hl7.org/fhir/json.html, Jul 2021.

\bibitem{kaul_rapidly_2020}
S.~Kaul, C.~Coleman, and D.~Gotz.
\newblock A rapidly deployed, interactive, online visualization system to
  support fatality management during the coronavirus disease 2019 ({COVID}-19)
  pandemic.
\newblock {\em Journal of the American Medical Informatics Association},
  27(12):1943--1948, Dec. 2020. doi: {{%
10\hspace{.1pt}\discretionary{.}{%
}{.}\hspace{.4pt}1093\discretionary{/}{%
}{/}jamia\discretionary{/}{%
}{/}ocaa146}}


\bibitem{ncats_n3c}
{NIH National Center for Advancing Translational Sciences}.
\newblock {National COVID Cohort Collaborative (N3C)}.
\newblock https://ncats.nih.gov/n3c, Jun 2021.

\bibitem{valuesets}
{NIH National Library of Medicine}.
\newblock {Value Set Authority Center}.
\newblock https://vsac.nlm.nih.gov/, Jul 2021.

\bibitem{athena}
{OHDSI}.
\newblock {ATHENA Standardized Vocabularies}.
\newblock
  https://www.ohdsi.org/analytic-tools/athena-standardized-vocabularies/, Jul
  2021.

\bibitem{plaisant_lifelines:_1998}
C.~Plaisant, R.~Mushlin, A.~Snyder, J.~Li, D.~Heller, and B.~Shneiderman.
\newblock {LifeLines}: using visualization to enhance navigation and analysis
  of patient records.
\newblock {\em Proceedings of the AMIA Symposium}, pp. 76--80, 1998.

\bibitem{preim_survey_2020}
B.~Preim and K.~Lawonn.
\newblock A {Survey} of {Visual} {Analytics} for {Public} {Health}.
\newblock {\em Computer Graphics Forum}, 39(1):543--580, 2020. doi: {{%
10\hspace{.1pt}\discretionary{.}{%
}{.}\hspace{.4pt}1111\discretionary{/}{%
}{/}cgf\hspace{.1pt}\discretionary{.}{%
}{.}\hspace{.4pt}13891}}


\bibitem{strausz_sleep_2021}
S.~Strausz, T.~Kiiskinen, M.~Broberg, S.~Ruotsalainen, J.~Koskela, A.~Bachour,
  FinnGen, A.~Palotie, T.~Palotie, S.~Ripatti, and H.~M. Ollila.
\newblock Sleep apnoea is a risk factor for severe {COVID}-19.
\newblock {\em BMJ Open Respiratory Research}, 8(1):e000845, Jan. 2021. doi:
  {{%
10\hspace{.1pt}\discretionary{.}{%
}{.}\hspace{.4pt}1136\discretionary{/}{%
}{/}bmjresp\discretionary{%
}{-}{-}2020\discretionary{%
}{-}{-}000845}}


\bibitem{thomas_illuminating_2005}
J.~Thomas and K.~Cook.
\newblock {\em Illuminating the {Path}: {The} {Research} and {Development}
  {Agenda} for {Visual} {Analytics}}.
\newblock National Visualization and Analytics Ctr, 2005.

\bibitem{torjesen2021covid}
I.~Torjesen.
\newblock Covid-19 will become endemic but with decreased potency over time,
  scientists believe.
\newblock {\em BMJ: British Medical Journal (Online)}, 372, 2021.

\bibitem{wang_role_2021}
H.~Wang, Z.~Yuan, M.~A. Pavel, S.~M. Jablonski, J.~Jablonski, R.~Hobson,
  S.~Valente, C.~B. Reddy, and S.~B. Hansen.
\newblock The role of high cholesterol in age-related {COVID19} lethality.
\newblock {\em bioRxiv}, p. 2020.05.09.086249, June 2021. doi: {{%
10\hspace{.1pt}\discretionary{.}{%
}{.}\hspace{.4pt}1101\discretionary{/}{%
}{/}2020\hspace{.1pt}\discretionary{.}{%
}{.}\hspace{.4pt}05\hspace{.1pt}\discretionary{.}{%
}{.}\hspace{.4pt}09\hspace{.1pt}\discretionary{.}{%
}{.}\hspace{.4pt}086249}}


\bibitem{west_innovative_2015}
V.~L. West, D.~Borland, and W.~E. Hammond.
\newblock Innovative information visualization of electronic health record
  data: a systematic review.
\newblock {\em Journal of the American Medical Informatics Association},
  22(2):330--339, Mar. 2015. doi: {{%
10\hspace{.1pt}\discretionary{.}{%
}{.}\hspace{.4pt}1136\discretionary{/}{%
}{/}amiajnl\discretionary{%
}{-}{-}2014\discretionary{%
}{-}{-}002955}}


\bibitem{wongsuphasawat_outflow:_2011}
K.~Wongsuphasawat and D.~Gotz.
\newblock Outflow: {Visualizing} {Patient} {Flow} by {Symptoms} and {Outcome}.
\newblock In {\em {IEEE} {VisWeek} {Workshop} on {Visual} {Analytics} in
  {Healthcare}}. Providence, Rhode Island, USA, 2011.

\bibitem{wongsuphasawat_lifeflow:_2011}
K.~Wongsuphasawat, J.~A. Guerra~Gómez, C.~Plaisant, T.~D. Wang,
  M.~Taieb-Maimon, and B.~Shneiderman.
\newblock {LifeFlow}: visualizing an overview of event sequences.
\newblock In {\em Proceedings of the 2011 annual conference on {Human} factors
  in computing systems}, {CHI} '11, pp. 1747--1756. ACM, New York, NY, USA,
  2011. doi: {{%
10\hspace{.1pt}\discretionary{.}{%
}{.}\hspace{.4pt}1145\discretionary{/}{%
}{/}1978942\hspace{.1pt}\discretionary{.}{%
}{.}\hspace{.4pt}1979196}}


\bibitem{zhang_dynamic_2019}
J.~Zhang, D.~Borland, W.~Wang, J.~Shrestha, and D.~Gotz.
\newblock Dynamic {Hierarchical} {Aggregation}, {Selection} {Bias} {Tracking},
  and {Detailed} {Subset} {Comparison} for {High}-{Dimensional} {Event}
  {Sequence} {Data}.
\newblock In {\em 2019 {IEEE} {Workshop} on {Visual} {Analytics} in
  {Healthcare} ({VAHC})}, pp. 56--57, Oct. 2019. doi: {{%
10\hspace{.1pt}\discretionary{.}{%
}{.}\hspace{.4pt}1109\discretionary{/}{%
}{/}VAHC47919\hspace{.1pt}\discretionary{.}{%
}{.}\hspace{.4pt}2019\hspace{.1pt}\discretionary{.}{%
}{.}\hspace{.4pt}8945029}}


\end{thebibliography}
